\documentclass[twocolumn,twoside,slac_two]{revtex4}
\usepackage{graphicx}
\usepackage{fancyhdr}
\begin{document}

\title{Weibel Instability Driven by Relativistic
Pair Jets: Particle Acceleration, Magnetic Field Generation, and
Emission}


\author{K.-I. Nishikawa}
\affiliation{National Space Science and Technology Center,
  Huntsville, AL 35805 USA}

\author{P. Hardee}
\affiliation{Department of Physics and Astronomy,
  University of Alabama,
  Tuscaloosa, AL 35487 USA}

\author{C. B. Hededal}
\affiliation{Niels Bohr Institute, Department of Astrophysics,
Juliane Maries Vej 30, 2100 K\o benhavn \O, Denmark}

\author{G. Richardson}
\affiliation{Department of Mechanical and Aerospace Engineering
  University of Alabama in Huntsville, AL 35899 USA}

\author{H. Sol}
\affiliation{LUTH, Observatore de Paris-Meudon, 5 place Jules Jansen
  92195 Meudon Cedex, France}

\author{R. Preece}
\affiliation{Department of Physics,
  University of Alabama in
  Huntsville, AL 35899 and National Space Science and Technology Center,
  Huntsville, AL 35805 USA}

\author{G. J. Fishman}
\affiliation{NASA-Marshall Space Flight Center, \\
National Space Science and Technology Center,
  Huntsville, AL 35805 USA}

\begin{abstract}
Shock acceleration is a ubiquitous phenomenon in astrophysical
plasmas. Plasma waves and their associated instabilities (e.g.,
Buneman, Weibel and other two-stream instabilities) created in
collisionless shocks are responsible for particle (electron,
positron, and ion) acceleration. Using a 3-D relativistic
electromagnetic particle (REMP) code, we have investigated particle
acceleration associated with a relativistic  jet front propagating
into an ambient plasma. We find that the growth times of the Weibel
instability in electron-positron jets are not affected by the
(electron-positron or electron-ion) ambient plasmas. However, the
amplitudes of generated local magnetic fields in the electron-ion
ambient plasma are significantly larger than those in the
electron-positron ambient plasma. The small scale magnetic field
structure generated by the Weibel instability is appropriate to the
generation of ``jitter'' radiation from deflected electrons
(positrons) as opposed to synchrotron radiation. The jitter
radiation resulting from small angle deflected electrons may be
important for understanding the complex time structure and spectral
evolution observed in gamma-ray bursts and other astrophysical
sources containing relativistic jets and relativistic collisionless
shocks.

\end{abstract}

\maketitle

\thispagestyle{fancy}


\section{INTRODUCTION}
Nonthermal radiation observed from astrophysical systems containing
relativistic jets and shocks, e.g., active galactic nuclei (AGNs),
gamma-ray bursts (GRBs), and Galactic microquasar systems usually
has power-law emission spectra. In most of these systems, the
emission is thought to be generated by accelerated electrons through
the synchrotron and/or inverse Compton mechanisms. Radiation from
these systems is observed in the radio through the gamma-ray region.
Radiation in optical and higher frequencies typically requires
particle re-acceleration in order to counter radiative losses.

Fermi acceleration is the mechanism usually assumed for the
acceleration of particles in astrophysical environments
characterized by a power-law spectrum. This mechanism for particle
acceleration relies on the shock jump conditions in relativistic
shocks [e.g., 1, 2]. Most astrophysical shocks are collisionless
since dissipation is dominated by wave-particle interactions rather
than particle-particle collisions. Diffusive shock acceleration
(DSA) relies on repeated scattering of charged particles by magnetic
irregularities (Alfv\'en waves) to confine the particles near the
shocks. However, particle acceleration near relativistic shocks
cannot be characterized as DSA because the propagation of
accelerated particles ahead of the shock cannot be described by
spatial diffusion. Anisotropies in the angular distribution of the
accelerated particles are large, and the diffusion approximation for
spatial transport does not apply [3].

\begin{figure*}[t]
\centering
\includegraphics[width=140mm]{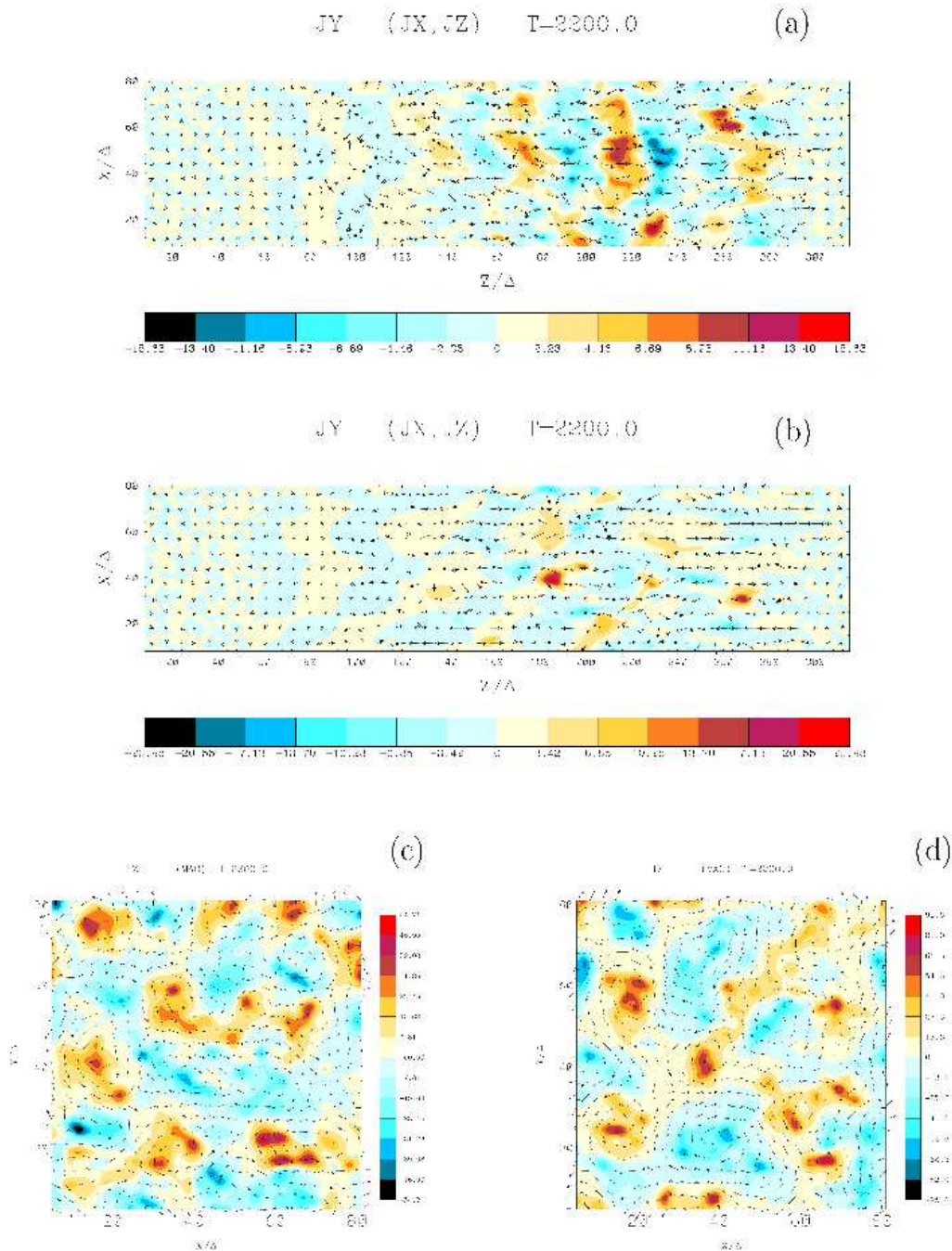}
\caption{2D images in the $x - z$ plane at $y = 43\Delta$ ((a) and
(b)) and in the $x - y$ plane at $z = 230\Delta$  a flat jet
injected into a unmagnetized ambient medium shown at $t =
28.8/\omega_{\rm pe}$. Colors indicate the $y$-component of the
current density ($J_{\rm y}$) ((a) and (b)) and $z$-component of the
current density ($J_{\rm z}$) ((c) and (d)) for electron-positron
((a) and (c)) and electron-ion ((b) and (d)) ambient plasmas. Their
peaks are (a) 15.6, (b) 23.9, (c) 54.7, and (d) 95.6. Arrows show
$J_{\rm z}, J_{\rm x}$ ((a) and (b)) and $B_{\rm x}, B_{\rm y}$ ((c)
and (d)).} \label{nishi-f1}
\end{figure*}

Particle-in-cell (PIC) simulations can shed light on the physical
mechanism of particle acceleration that occurs in the complicated
dynamics within relativistic shocks.  Recent PIC simulations using
injected relativistic electron-ion jets show that acceleration
occurs within the downstream jet, rather than by the scattering of
particles back and forth across the shock as in Fermi acceleration
[4-10]. In general, these independent simulations have confirmed
that relativistic jets excite the Weibel instability [11]. The
Weibel instability generates current filaments with associated

\begin{figure*}[th]
\includegraphics[width=175mm]{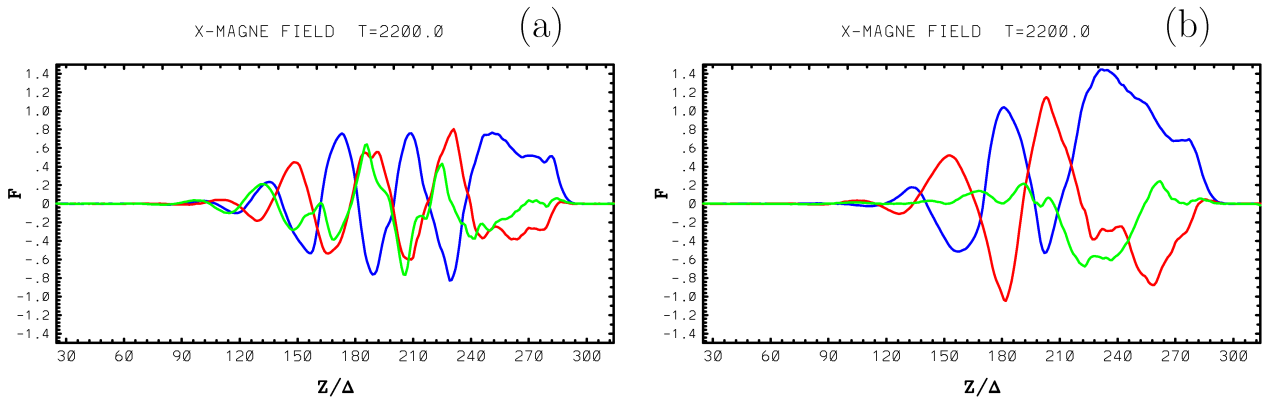}
\caption{One-dimensional cuts along the $z$-direction ($25 \leq
z/\Delta \leq 154$) of a flat jet. Shown are the $x$-components of
the magnetic field shown at $t = 238.8/\omega_{\rm pe}$ for
unmagnetized electron-positron (a) and  electron-ion (b) ambient
plasmas. Cuts are taken at $x/\Delta = 38$ and $y/\Delta = 33
(blue), 43 (red), 53 (green)$ and separated by about an electron
skin depth.} \label{nishi-f2}
\end{figure*}

\noindent
magnetic fields [12], and accelerates electrons [4-10].

In this paper we present new simulation results of particle
acceleration and magnetic field generation in relativistic
electron-positron shocks using 3-D relativistic electromagnetic
particle-in-cell (REMP) simulations. We have performed two
simulations in which an electro-positron jet is injected into two
different ambient plasmas (electron-positron and electron-ion). In
our new simulations, the growth rate of the Weibel instability and
its evolution with different ambient plasmas have been studied
without an initial ambient magnetic field.

\section{SIMULATION SETUP AND RESULTS}

Two simulations were performed using an $85 \times 85 \times 320$
grid with a total of 180 million particles (27
particles$/$cell$/$species for the ambient plasma) and an electron
skin depth, $\lambda_{\rm ce} = c/\omega_{\rm pe} = 9.6\Delta$,
where $\omega_{\rm pe} = (4\pi e^{2}n_{\rm e}/m_{\rm e})^{1/2}$ is
the electron plasma frequency and $\Delta$ is the grid size [5, 8].
% In all
%simulations jets are injected at $z = 25\Delta$ in the positive $z$
%direction. Radiating boundary conditions were
%used on the planes at $z =0,z_{\rm max}$. Periodic boundary
%conditions were used on all other boundaries (Buneman 1993).
%

The particle number density of the jet is $0.741n_{\rm b}$, where
$n_{\rm b}$ is the density of ambient (background) electrons. The
average jet velocities in the two simulations are $v_{\rm j} =
0.9798c$ corresponding to Lorentz factors of 5 (2.0 MeV). The jets
are cold ($v^{\rm e}_{\rm j, th} = v^{\rm p}_{\rm j, th} = 0.01c$)
in the rest frame of the ambient plasma where $c$ is the speed of
light. Electron-positron plasmas have mass ratio $m_{\rm p}/m_{\rm
e} = 1$. The mass ratio of ion and electron is $m_{\rm i}/m_{\rm e}
= 20$. The electron (positron) and ion thermal velocities in the
ambient plasmas are $v^{\rm e}_{\rm th} = v^{\rm p}_{\rm th} = 0.1c$
and $v^{\rm i}_{\rm th} = 0.022c$, respectively. The time step
$\Delta t = 0.013/\omega_{\rm pe}$.

% The jet makes contact with the ambient plasma at a 2D
%interface spanning the computational domain. Here only the dynamics
%of the propagating jet head and shock region is studied. Effectively
%we study a small uniform portion of a much larger shock. This
%simulation system is different from simulations performed using
%counter-streaming equal number density particles spanning the
%computational domain in the transverse direction. The important
%differences between this type of simulation and previous
%counter-streaming simulations is that the evolution of the Weibel
%instability is examined in a more realistic spatial way including
%the motion of the jet head, and we can have different number
%densities in beam and ambient medium.

 The electrons are deflected by the transverse magnetic
fields ($B_{\rm x}, B_{\rm y}$) via the Lorentz force: $-e({\bf v}
\times {\bf B})$, generated by current filaments ($J_{\rm z}$),
which in turn enhance the transverse magnetic fields [5, 6, 8]. The
complicated filamented structures resulting from the Weibel
instability have diameters on the order of the electron skin depth
($\lambda_{\rm ce} = 9.6\Delta$). This is in good agreement with the
prediction of $\lambda \approx 2^{1/4}c\gamma_{\rm
th}^{1/2}/\omega_{\rm pe} \approx 1.188\lambda_{\rm ce} =
11.4\Delta$ \cite{medv99}. Here, $\gamma_{\rm th} \sim 1$ is a
thermal Lorentz factor. At the earlier time smaller current
filaments are generated. However, in the electron-positron jets
because of larger growth rates, the current filaments are coalesced
in the transverse direction in the nonlinear stage as shown by
Nishikawa et al. [13, 14]. The longitudinal current ($J_{\rm z}$) in
the electron-positron jet shows significantly more transverse
variation than in the electron-ion jets [13, 14].

Current filaments resulting from development of the Weibel
instability behind the jet front are shown in Fig. 1 at time $t =
28.8/\omega_{\rm pe}$ for unmagnetized electron-positron ((a) and
(c)) and  electron-ion ((b) and (d)) ambient plasmas. Figures 1a and
1b show the $y$-component of the current density in the $z - x $
plane at $y = 43\Delta$. The maximum values of $J_{\rm y}$ are (a)
15.6 and (b) 23.9, respectively. The arrows show the perturbed
currents ($J_{\rm z}, J_{\rm x}$). Figures 1c and 1d show the
$z$-component of the current density in the $x - y$ plane at $z =
230\Delta$. The arrows show the magnetic fields ($B_{\rm x}, B_{\rm
y}$) created by  $J_{\rm z}$ and their maximum values are (c) 54.7
and (d) 95.6, respectively. Based on these figures the maximum
values of the generated current densities (magnetic fields) are
larger in the electron-ion ambient plasmas. The ambient ions are
responsible for the larger maximum current densities.

It should be noted that as shown in the previous work for the
different jet types ((a) electron-positron and (b) electron-ion)
injected into ambient plasmas with the jet composition, the maximum
current densities have different structures [13, 14]. Current
filaments resulting from development of the Weibel 

\begin{figure*}[t]
\includegraphics[width=175mm]{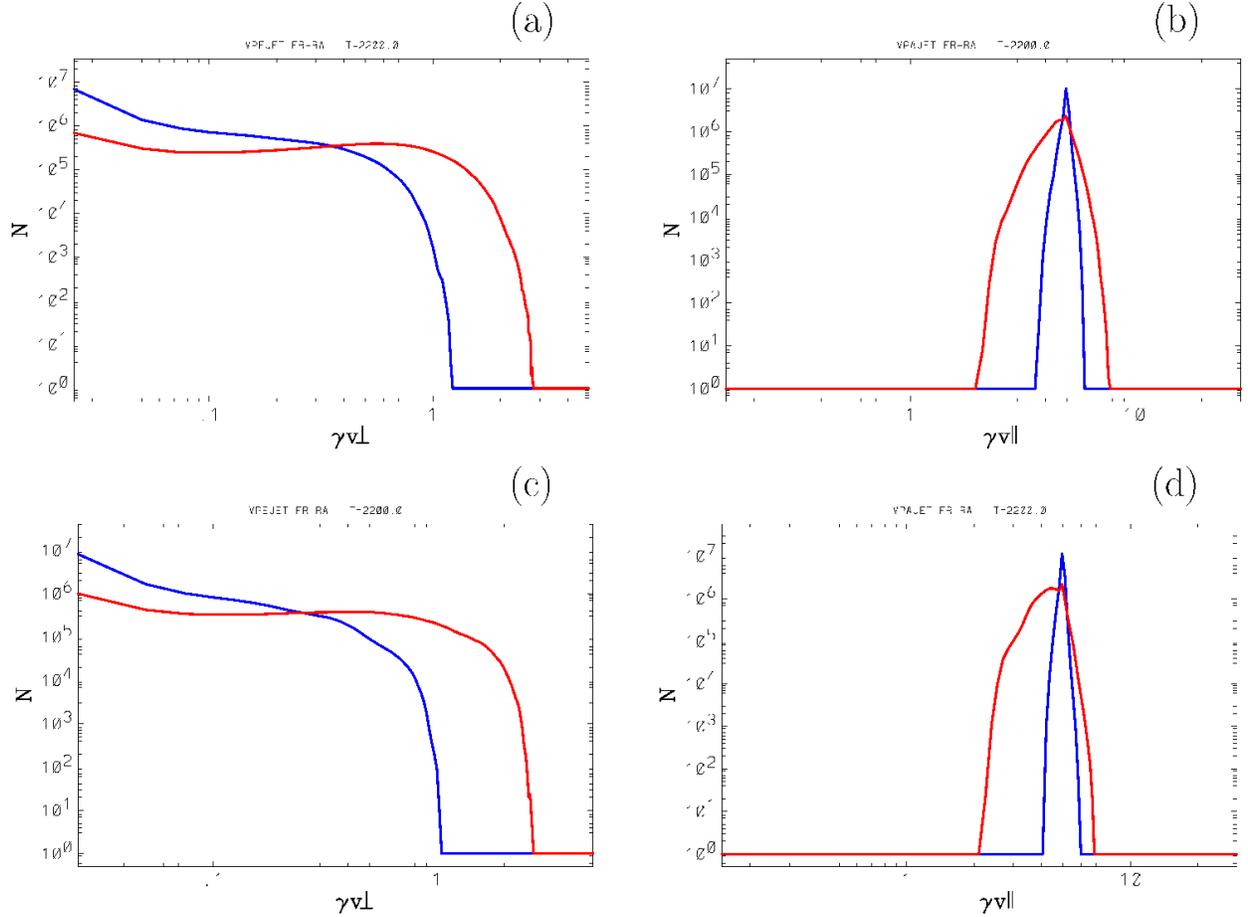}
\caption{Velocity distributions of jet electrons two cases as Fig. 1
at $t = 28.8/\omega_{\rm pe}$ ((a)and (b): electron-positron ambient
plasma, (c) and (d) electron-ion ambient plasma. Jet electrons are
binned as a function of $\gamma v_{\perp}$ ((a) and (c)) and $\gamma
v_{\parallel}$ ((b) and (d)), and $\gamma v_{\perp}$, where $\gamma
= (1 -(v^{2}_{\parallel} +v^{2}_{\perp})/c^{2})^{-1/2}$. The blue
and red curves show the distributions of injected and shocked jet
electrons.} \label{nishi-f3}
\end{figure*}

instability
behind the jet front (at $z = 230\Delta$) are obtained at time $t =
28.8/\omega_{\rm pe}$ for two different cases with unmagnetized
ambient plasmas. The maximum values of $J_{\rm z}$ are (a) $\pm
54.7$ (as same as Fig. 1a), (b)$ -123.6$, respectively. The
electro-positron jets show electron (negative) and positron
(positive) current filamentations, since both species contribute to
the Weibel instability. On the other hand, at this simulation time
mainly electron jets generate (negative) current filamentations. In
the electron-ion jet at this time only electrons are contributing in
exciting the Weibel instability, electron current channels are
dominate, therefore $J_{\rm z}$ has the negative value.

The differences in the generated current densities ($J_{\rm z}$)
between the different ambient plasmas are seen more clearly in the
$x$-component of the generated 
magnetic fields as shown in Fig. 2. The amplitudes of
$B_{\rm x}$ in the electron-ion ambient plasma (b) are much larger
than those in the electron-positron ambient plasma (a). The
distances between peaks in (b) are clearly larger than those in (a).
This seems to come from the fact that the ambient ions are affecting
the growth of the Weibel instability. These differences in the
amplitudes and structures will affect the emission, which needs to
be calculated from an ensemble of all particles (electrons and
positrons). Since magnetic fields are generated by the Weibel
instability ``jitter'' radiation need to be considered by the
deflected electrons (positrons) [15, 16, 17]. In order to examine
the saturated magnetic fields we need much longer simulations, which
are in progress at the present time.

The Weibel instability is aperiodic, i.e., $\omega_{\rm real} \sim
0$ (convective) [12]. Thus, it can be saturated only by nonlinear
effects and not by kinetic effects, such as collisionless damping or
resonance broadening. Hence the magnetic field can be amplified to
very high values. This characteristic is seen in Fig. 2b. The time
evolution of the $B_{\rm x}$ shows that its amplitudes grow as the
jet propagates without any significant oscillations in the frame of
the jet.

Simulation results show that smaller scale current filaments appear
immediately behind the jet front. Smaller filaments merge into
larger filaments behind the jet front in the non-linear stage. This
phenomenon is seen in Fig. 2. In particular Figure 2b shows a large
and long wavelength just behind the jet front.

It should be noted that a long simulation with electron-ion jet with
$\gamma = 15$ show that the electron Weibel instability switches to
the ion Weibel instability [9]. The ion current channels with the
ion skin depth contribute to accelerating electrons [4, 7, 9].

The acceleration of electrons has been reported in previous work
[4-10, 13, 14]. Figure 3 shows that the cold jet electrons are
accelerated and decelerated. As expected, at this time jet electrons
in the electron-positron ambient plasma (b) are thermalized more
strongly than those in the electron-ion ambient plasma (d). The blue
curves in Figs. 3b and 3d are close to the initial distribution of
injected jet electrons (half of jet electrons). We also see that the
kinetic energy (parallel velocity $v_{\parallel} \approx v_{\rm j}$)
of the jet electrons is transferred to the perpendicular velocity
via the electric and magnetic fields generated by the Weibel
instability [11, 12]. The strongest transverse acceleration of jet
electrons accompanies the strongest deceleration of electron flow
and occurs between $z/\Delta = 210 - 240$.  The transverse velocity
distribution of jet in the electron-positron ambient plasma (at the
peak value) is slightly more accelerated than that in the
electron-ion ambient plasma as shown in the red curves in Figs. 3a
and 3c. The strongest acceleration takes place around the maximum
amplitude of perturbations due to the Weibel instability at
$z/\Delta \sim 230$ as seen in Figs 1a and 1b.

\section{SUMMARY AND DISCUSSION}
We have performed self-consistent, three-dimensional relativistic
particle simulations of relativistic electron-positron jets
propagating into unmagnetized electron-positron and electron-ion
ambient plasmas. The main acceleration of electrons takes place in
the region behind the shock front [4-5, 7-10, 13, 14, 17]. Processes
in the relativistic collisionless shock are dominated by structures
produced by the Weibel instability. This instability is excited in
the downstream region behind the jet head, where electron density
perturbations lead to the formation of current filaments. The
nonuniform electric field and magnetic field structures associated
with these current filaments decelerate the jet electrons and
positrons, while accelerating the ambient electrons and positrons,
and accelerating (heating) the jet and ambient electrons and
positrons in the transverse direction.

The effects of the different ambient plasmas have been investigated
in this report. In spite of the local larger current filaments
generated in the electron-ion ambient plasma, the total jet electron
acceleration is smaller than that in the electron-positron ambient
plasma.

The growth rates depend on the Lorentz factors of the jet as
suggested by the theory [12]. $E$-fold time is written as $\tau
\simeq \sqrt{\gamma_{\rm sh}}/\omega_{\rm pe}$, where $\gamma_{\rm
sh}$ is the Lorentz factor of the shock. The simulation results show
that the growth time becomes larger with the larger Lorentz factor
[18].

Other simulations with different skin depths and plasma frequencies
confirm that both simulations have enough resolution and the
electron Weibel instability is characterized by the electron skin
depth [8].

 An additional simulation in which an
electron-ion jet is injected into an ambient plasma with
perpendicular magnetic field shows magnetic reconnection due to the
generation of an antiparallel magnetic field generated by bending of
jet electron trajectories, consequently jet electrons are subject to
strong non-thermal acceleration [10].
% (Hededal and Nishikawa 2004).

The generation of magnetic fields both with and without an initial
magnetic field suggests that emission in GRB afterglows and
Crab-like pulsar winds could be either synchrotron or jitter
emission [15, 16, 17].  The filament sizes appear to be smaller than
can produce observable variations in intensity structure. However,
this small size can mean that the deflection angle, $\alpha \sim e
B_{\perp} \lambda_{\rm B}/\gamma m_e c^{2}$, of particles by Weibel
filaments is smaller than the radiation beaming angle, $\Delta\theta
\sim 1/\gamma$ [12].  Here $\lambda_{\rm B} \sim \lambda_{\rm ce}$,
$e B_{\perp}/m_e c < \Omega_{\rm e}$, and the ratio $\delta \sim
\alpha/\Delta\theta < \Omega_{\rm e}/\omega_{\rm pe}$ will be less
than one when the cyclotron frequency is less than the plasma
frequency.  Thus, when ambient magnetic fields are moderate, i.e.,
the cyclotron frequency is less than the plasma frequency and
$\delta < 1$, the emission may correspond to jitter rather than
synchrotron radiation [15, 16, 17].

The fundamental characteristics of relativistic shocks are essential
for a proper understanding of the prompt gamma-ray and afterglow
emission in gamma-ray bursts, and also to an understanding of the
particle reacceleration processes and emission from the shocked
regions in relativistic AGN jets.  Since the shock dynamics is
complex and subtle, more comprehensive studies are required to
better understand the acceleration of electrons, the generation of
magnetic fields and the associated emission. This further study will
provide insight into basic relativistic collisionless shock
characteristics needed to provide a firm physical basis for modeling
the emission from shocks in relativistic flows.

\bigskip % extra skip inserted
\begin{acknowledgments}
K. Nishikawa is a NRC Senior Research Fellow at NASA Marshall Space
Flight Center. This research (K.N.) is partially supported by the
National Science Foundation awards ATM-0100997, and INT-9981508. P.
Hardee acknowledges partial support by a National Space Science and
Technology (NSSTC/NASA) award.  The simulations have been performed
on IBM p690 at the National Center for Supercomputing Applications
(NCSA) which is supported by the National Science Foundation.

\end{acknowledgments}

\bigskip % extra skip inserted

\end{document}